\documentclass[11pt]{article}
\usepackage{setspace} 
\usepackage{amsmath}

\usepackage{graphicx}

\usepackage[round,numbers,sort&compress]{natbib} 
\title{Electrophoresis of positioned nucleosomes}

\author{Martin Castelnovo\thanks{
           Corresponding author.} \\
	Laboratoire Joliot-Curie et Laboratoire de Physique, \\
	Ecole Normale Sup\'erieure de Lyon, \\
        46 All\'ee d'Italie, 69364 Lyon Cedex 07, France 
	\and S\'ebastian Grauwin \\
	Laboratoire Joliot-Curie et Laboratoire de Physique, \\
	Ecole Normale Sup\'erieure de Lyon,  \\
        46 All\'ee d'Italie, 69364 Lyon Cedex 07, France}

\date{}

\begin{document}

\maketitle

\abstract{We present in this paper an original approach to compute the electrophoretic mobility of rigid nucleo-protein complexes like nucleosomes. This model allows to address theoretically the influence of complex position along DNA, as well as wrapped length of  DNA on the electrophoretic mobility of the complex. The predictions of the model are in qualitative agreement with experimental results on mononucleosomes assembled on short DNA fragments ($<400bp$). Influence of additional experimental parameters like gel concentration, ionic strength, effective charges is also discussed in the framework of the model, and is found to be qualitatively consistent with experiments when available.  Based on the present model, we propose a simple semi-empirical formula describing positioning of nucleosomes as seen through electrophoresis.

}

\clearpage

\section*{Introduction}
Electrophoresis is one of the most powerful and widely used technique in modern molecular biology in order to address various properties of biological samples: molecular weight and size determination (DNA, proteins ...), mapping of particular protein binding sites on DNA (enzyme footprinting), effective charges (protein charge ladders) \citep{Lodish_etal99}. In most applications aforementioned, there is a need of some a calibrated sample, the so-called ``ladder'', in order to quantify the results of any electrophoresis experiments. This allows to use these techniques without the precise knowledge of physical mechanisms underlying electrophoresis separation. Nevertheless, by processing this way, one misses additional informations that are not brought by the comparison with the ladder. This is the case for example for nucleo-protein complexes like mononucleosomes. The nucleosome is the first degree of organization of DNA within the chromatin of eukaryotes. It is made from the complexation of roughly 147 bp of DNA with an octamer of histone proteins \citep{Luger_etal97}. It has been shown \textit{indirectly} that electrophoretic mobility of mononucleosomes depends on its positioning along DNA \citep{Pennings97}. Now this property is widely used to detect qualitatively nucleosome repositioning due either to thermal fluctuations or to the action of remodeling factors \citep{Flaus_etal03,Angelov_etal04}. But taken alone, these kind of experiments do just indicate that a change occured either on the conformation of nucleosome, and/or on its charge distribution, since electrophoresis of colloidal particles is mostly sensitive to these two intrinsic properties. No quantitative conclusions can be reached with respect to the precise position of nucleosome along DNA. Physical modeling of electrophoresis might help extracting this information from the experiments.

In the early works of Pennings and collaborators about position-dependent electrophoretic mobility of nucleosomes \citep{Pennings97}, datas were interpreted using similar results obtained on short bent oligonucleotides \citep{Koo_etal88,Drak_etal91}: the mobility of such molecules in a gel is strongly dependent on both position and angle of the bent, with same qualitative trends. Apart from the similarity of experimental results, the systems are not expected to behave exactly the same, due to the large size of the nucleosome core, roughly 10 nm in diameter, which is not present in the bent oligonucleotides experiments. But it is quite likely that position selectivity arises from the same physical mechanisms, still to be discovered. On the theoretical side, reptation models have been shown to explain qualitatively some features about bent-DNA but the position-dependent mobility cannot  be obtained in a quantitative way \citep{Levene_etal89,Drak_etal91}. Again, the large size of the nucleosome core renders the reptation mechanism difficult to apply in our case, at least for explaining this position-dependence of mobility. Therefore there is a real specifity of large DNA-protein complexes with respect to their mobility in pure solution or in gels, as compared to naked bent-DNA. In the present study, we will focus on the nucleosomal case. The methods developped in this context are currently applied to analyze position-dependent mobility of bent DNA in a separate study \citep{Castelnovo_06}. 

Precise computation and description of electrophoretic mobility is a formidable task as soon as non-trivial geometries are considered. Indeed one need to solve simultaneously equations describing electrostatic potential, flow profile and ionic species distribution \citep{Allison_etal97}. Moreover, any realistic model should take into account the effect of sieving medium in which electric migration is performed. We propose in this work a general method to evaluate electrophoretic mobility of a rigid nucleo-protein complex in pure buffer or in gels through effective continuous electro-hydrodynamic description: mimicking the conformation of nucleosome by a set of charged beads of appropriate size and charge, we calculate total electrophoretic mobility similarly to the way friction coefficients of proteins are evaluated using beads models mapping the protein conformation \citep{Allison_etal98,delaTorre_etal81}. This approach has been already applied to study theoretically the influence of different charge distribution on the electrophoretic mobility of polyampholytes \citep{Long_etal98}. Moreover, it will be shown that in order to reproduce quantitatively the experimental results specific gel features will have also to be taken into account in the model. Taking all these theoretical ingredients together allows then to investigate the influence of different physical factors like gel concentration, buffer ionic strength, bead-complex conformation on the electrophoretic mobility.

In order to illustrate the benefits of such an approach, we address two original questions in the context of mononucleosomes characterization: (\textit{i}) is the position-dependent electrophoretic mobility to be seen in pure buffer, without any sieving medium, corresponding to the case of capillary electrophoresis, and (\textit{ii}) what is the influence of nucleosome geometry like the amount of DNA length wrapped around the histone core within the nucleosome on its electrophoretic mobility? The first question allows to address the role of the gel in position-dependent mobility. In the context of the second question, the non-canonical conformations of a mononucleosome are supposed to mimick different incomplete states of nucleosomes. As an example, it is known that the four different histones (H2A,H2B,H3,H4) found in canonical nucleosomes are arranged into an octamer. Two different type of partial association of histones leading to DNA-histones complexes can also be found in solution: H3-H4 tetramers, and hexamers made of one H3-H4 tetramer and one H2A-H2B dimer. These are characterized by different amount of DNA wrapped around the protein core. Experimentally, they have different electrophoretic mobilities. Another recent example of interest is the case of nucleosomes made of histone variants, which are found in the chromatin at some specific locations along the genome where a strong regulation of gene expression occurs (either repression or activation) \citep{Henikoff_etal05}.
In the case of the variant H2A.Bbd, it is believed that DNA wrapped length in the nucleosome variant is of order 120 base pairs instead of the canonical 147 base pairs \citep{Bao_etal04}. Within our model, it is possible to evaluate the difference in electrophoretic mobility between canonical and variant nucleosomes for the same DNA length.

The paper is organized as follows. In the next section, we describe first the general formalism to compute the electrophoretic mobility of a coarse-grained nucleosome model through continuous electro-hydrodynamics, and then the way to include specific gel effects in this model. Numerical results of such models are then presented. Finally applications and limitations of the present work are discussed.

\section*{Model}

\subsection*{General formalism}
\label{model section}In the present work, we propose a coarse-grained model of a mononucleosome: its shape and total charge are approximated by a rigid set $\{i\}$ of non-overlapping charged beads of radii $\sigma_i$ and net charges $z_i$, hereafter denoted as the bead-complex (cf fig.~\ref{figure1} a). The net steady state  motion of such an object under an external electric field $\bf{E}$ in a buffer of ionic strength $I$ and viscosity $\eta$ is due to the balance between electrostatic and hydrodynamic forces. The rigidity assumption amounts to neglect conformation fluctuations of the whole complex and especially of DNA arms. This is justified for the latter as long as the arms are shorter than a persistence length, \textit{i.e.} the thermal rigidity length scale, which sets the upper limit of total DNA length (wrapped length and arms) to roughly 400 base pairs. The neglect of conformation fluctuations of the complex is associated to the tight wrapping of DNA around nucleosomes. The influence of bead-complex opening angle fluctuations is then addressed within our model by computing the mobility for various rigid conformations.

A naive statement for a rigid object like the bead-complex would be that the electrostatic forces are purely driving the motion, while the hydrodynamic forces purely exert drags, just like in any sedimentation or centrifugation experiments. However, it is well-known that electrostatic forces contribute as well to the net hydrodynamic drag due to the presence of counterions going in reverse direction of motion and therefore exerting an additional drag. This is the very presence of co and counterions that makes the problem of calculating exactly the electrophoretic mobility a tedious task: full solution of the problem would require to solve simultaneously Poisson equation for the electrostatic potential, Navier-Stokes equation for the flows and ion transport equation for the spatial distributions of ions \citep{Allison_etal97}.

Under certain range of parameters, it is however possible to obtain a simple closed formula by using several assumptions. The first one is that the distributions of co and counterions around the bead-complex are equilibrium distributions. This amounts to neglect the so-called ion-relaxation effect which is important mainly for highly charged objects and high electric field \citep{Allison_etal97}. The main consequence of this assumption is that electrostatics is now described by classical Poisson-Boltzmann equation. The second assumption is that the Debye-Huckel linear approximation for the electrostatic potential is valid, \textit{i.e.} bead-complex is not highly charged. Finally, we assume that the electric field driving the motion of the bead-complex is small enough such that orientation and polarization effects are negligible. The validity of these assumptions with respect to realistic systems is discussed in section Applications and Limitations

Due to the linearity of Navier-Stokes equation at low Reynolds number as considered in this work, each bead subjected to a force contributes linearly to the flow field at any given point through hydrodynamic interactions. Following Long \textit{et al.} \citep{Long_etal98}, we identify two different types of force on each bead, that generate different hydrodynamic contributions. The first type of force $\{F_i\}$ is due to the rigid physical connection between neighbouring beads, and it is still present when electrostatic interactions are switched off. The associated long-range hydrodynamic interaction is accurately described by Rotne-Prager tensor  \citep{Yamakawa70,Rotne_etal69}, which is the first finite-volume correction to Oseen tensor associated to point-like forces
\begin{eqnarray}
{\bf T}^{RP}_{ij} & = & \left(1+\frac{\sigma_i^2+\sigma_j^2}{6}\nabla^2 \right){\bf T}^{O}_{ij}\\
{\bf T}^{O}_{ij} & = & \frac{1}{8\pi\eta r_{ij}}\left({\bf I}+\frac{{\bf r}_{ij}{\bf r}_{ij}}{r_{ij}^2}\right)
\end{eqnarray}
where $r_{ij}$ is the distance between centers of beads $i$ and $j$, and ${\bf I}$ is the identity tensor. 

The other type of force acting on each bead is electrostatic through the external electric field. Within Debye-Huckel approach, this generates a screened flow profile due to the presence of co and counterions in the solution \citep{Long_etal01,Allison_etal00,Huckel24,Allison_06}. The tensor to be used to describe this flow is obtained similarly to Rotne-Prager tensor
\begin{eqnarray}
\label{RP screened1}{\bf T}^{RPel}_{ij} & = & \left(1+\frac{\sigma_i^2+\sigma_j^2}{6}\nabla^2 \right){\bf T}^{el}_{ij}\\
{\bf T}^{el}_{ij} & = & \frac{e^{-\kappa_D r_{ij}}}{4\pi\eta r_{ij}}\bigg[\big(1+\frac{1}{\kappa_D r_{ij}}+\frac{1}{(\kappa_D r_{ij})^2}\big){\bf I}-\big(\frac{1}{3}+\frac{1}{\kappa_D r_{ij}}+\frac{1}{(\kappa_D r_{ij})^2}\big)\frac{3{\bf r}_{ij}{\bf r}_{ij}}{r_{ij}^2}\bigg]\nonumber\\
 & & -\frac{1}{4\pi\eta\kappa_D^2r_{ij}^3}\left[{\bf I}-\frac{3{\bf r}_{ij}{\bf r}_{ij}}{r_{ij}^2}\right]\label{RP screened2}
\end{eqnarray}
The Debye-Huckel screening length $\kappa_D^{-1}$ scales with ionic strength of the buffer $I$ like $\kappa_D^{-1} \sim I^{-1/2}$.

For pure translation motion, the velocity of each bead is therefore given by
\begin{eqnarray}
\label{v}{\bf v}_i & = & {\bf v}_i^0+\sum_{j\neq i}({\bf T}^{RP}_{ij}.{\bf F_j}+{\bf T}^{RPel}_{ij}.z_j {\bf E})\\
\label{v0}{\bf v}_i^0 & = & {\bf v}_i^{0,neutral}+\mu_i^0 {\bf E}\\
\label{F}{\bf F}_i & = & \xi_i^0 {\bf v}_i^{0,neutral}
\end{eqnarray}
in term of the velocity ${\bf v}_i^0$ due to each pure force field separately. The friction coefficient of a single bead is $\xi_i^0=6\pi\eta \sigma_i$, and the electrophoretic mobility of each bead regardless of the presence of the others is simply $\mu_i^0=\frac{z_i}{6\pi\eta \sigma_i}$.
Equations ~\ref{v} to ~\ref{F} can be cast into a single equation
\begin{equation}
\sum_j {\bf T}^{RP}_{ij}.{\bf F}_j=(\mu-\mu_i^0-\sum_{j\neq i}{\bf T}_{ij}^{RPel}.z_j){\bf E}
\end{equation}
where all beads have the same velocity ${\bf v}_i\equiv{\bf V}=\mu {\bf E}$ for a steady motion, and diagonal terms in the Rotne-Prager tensors are defined as ${\bf T}^{RP}_{ii}=1/\xi_i^0$. In the case of screened hydrodynamic interactions, the diagonal term is given in the Appendix A \citep{Long_etal01,Russel_etal89,Stigter00}. Notice that this term is also the inverse of isolated bead friction coefficient.

Using the fact that the ${\bf F}_i$'s are internal forces, the final result for the electrophoretic mobility for a given orientation of the bead-complex is
\begin{equation}
\label{mu principal}
\mu = \frac{\sum_{ijk}T_{||,ij}^{-1}T_{||,jk}^{el}z_k}{\sum_{ij}T_{||,ij}^{-1}}
\end{equation}
where the notation of tensors ${\bf T}^{RP}_{ij}$ and ${\bf T}^{RPel}_{ij}$  has been respectively simplified to ${\bf T}_{ij}$ and ${\bf T}^{el}_{ij}$. The index ``$||$'' means that tensors have been projected along the electric field direction. Notice also that $T^{-1}_{||,ij}$ is the inverse tensor of $T_{||,ij}$, such that $\sum_{j}T_{||,ij}T^{-1}_{||,jk}=\delta_{ik} $. This formal result was already obtained by Long \textit{et al.} in their discussion of polyampholyte dynamics \citep{Long_etal98}. However, they were mainly interested in the influence of different charge distributions for average conformations. In the present work, the conformation is fixed due to the assumed rigidity of the bead-complex, and therefore Eq.~\ref{mu principal} can \textit{directly} and \textit{explicitely} be used to calculate the electrophoretic mobility. This has the further advantage to keep track of bead-complex orientation with respect to the electric field $\varphi$. Results of the numerical calculation for particular geometries of the bead-complex in the case of mononucleosome are provided and discussed in the next sections. 

\subsection*{Specific gel effects}
\label{specific gel effects}Up to this point, no effect of sieving medium, \textit{i.e.} the polymeric gel (polyacrilamide, agarose...), has been taken into account. In this section, we describe three main effects due to the gel, and how they can be incorporated into the original model: hydrodynamic flow screening,  constrained orientation of bead-complex in the gel and trapping.

\textbf{Hydrodynamic screening} -- The first effect of the gel on the migration of bead-complex is to screen hydrodynamic flow \citep{Long_etal01}, as was originally proposed by Brinkman to describe hydrodynamics in porous media \citep{Brinkman47}. This effect is straightforwardly incorporated in the original continuous electro-hydrodynamics. Following  Long and Ajdari \citep{Long_etal01}, tensors describing screened hydrodynamic flow either due to electrophoretic motion or to neutral migration in a porous medium are identical. In the latter case, it is given by Rotne-Prager tensor in Eqs.~\ref{RP screened1}, the Debye screening length being replaced by the gel screening length $\kappa_g^{-1}=(\xi_g c_g)^{-1/2}$, where the gel is represented as a collection of beads of friction coefficients $\xi_g$ and concentration $c_g$. With this modification of long-ranged Rotne-Prager tensor into a short-range one, the electrophoretic mobility of the bead complex in a gel can be calculated according to Eq ~\ref{mu principal}.

\begin{figure}
   \begin{center}
      \includegraphics*[width=14cm]{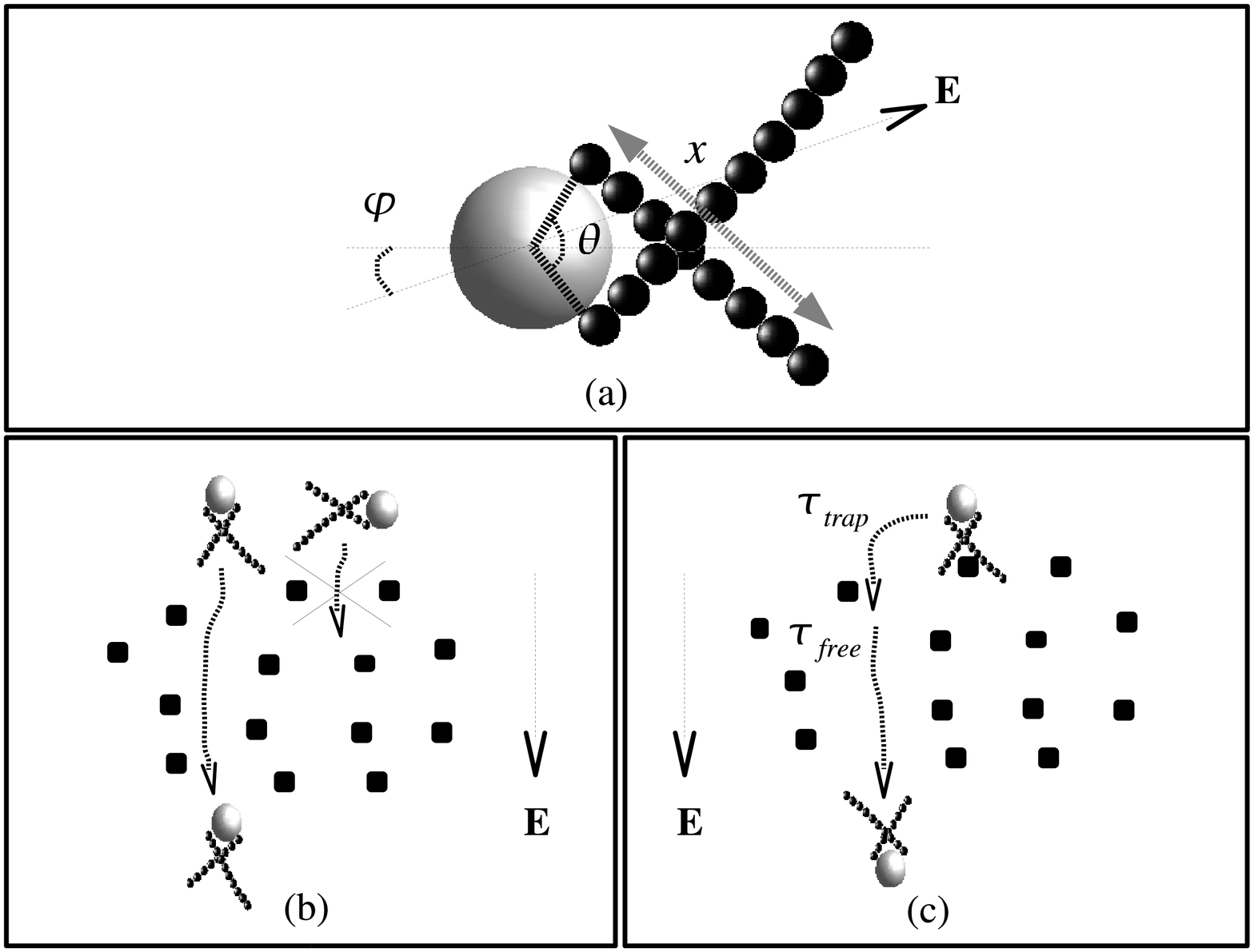}
      \caption{}
      \label{figure1}
   \end{center}
\end{figure}

\textbf{Orientation} -- The second important effect of the gel is to constrain the orientation of the bead-complex during its migration, see figure ~\ref{figure1}b. Indeed for an anisotropic object like a mononucleosome with finite length DNA arms, the migration is enhanced if the size of the complex in the direction perpendicular to the electric field is smaller than gel pore size, while it is strongly reduced in the reverse situation. Using the continuous electro-hydrodynamic model presented in the previous section,  this effect can be taken into account by constraining the range of orientation angle while performing the orientation average. This will be discussed more precisely in the next section.

\textbf{Trapping} -- Finally, the migration of nucleosomes within a gel is strongly influenced by trapping events: since there is a finite bending angle between DNA arms leaving the core of the nucleosome, this bent or kink might be trapped transiently through a collision with gel fibers (cf fig. ~\ref{figure1} c), just like long naked DNAs are known to hook in U-shape around gel fibers for high electric fields electrophoresis \citep{Viovy00}. The untrapping process in the case of naked DNA is thought to occur like a rope on a pulley. In the present case of rigid bead-complex, the escape from trapped configuration is mainly achieved by rigid rotation around the gel fiber. The overall average motion of the complex is then described by alternation of two states: \textit{(i)} a uniform steady motion in the free volume of the gel with a pure buffer velocity ${\bf v}$, corrected by the hydrodynamic screening effect of the gel previously mentioned, during an average time $\tau_{free}$, and \textit{(ii)} trapping/untrapping event of vanishing net velocity, during an average time $\tau_{trap}$. As a result, the average velocity $V$ in the direction of the electric field is given by
\begin{equation}
\label{trap}V=\frac{v}{1+\frac{\tau_{trap}}{\tau_{free}}}
\end{equation}
Similar formula has been used to describe the motion of long naked DNA in gel when trapping events are mainly determining the overall dynamics \citep{Dorfman_etal04,Popelka_etal99}.

The average time during free motion is estimated by the mean-free path of the bead-complex $l_{MFP}\sim 1/(\pi d^2c_g)$, with $d$ the diameter of gel fiber and $c_g$ its concentration. Therefore the estimate of $\tau_{free}$ is 
\begin{equation}
\tau_{free} \sim \frac{\pi d^2 c_g}{v}
\end{equation}

Although collision scenario is not precisely known during trapping events, one might anticipate that the longest time (which is the relevant time for the mobility calculation) is associated with the rotation of the complex around gel fibers. This motion is driven by the electrostatic torque $\Gamma_{el}(\varphi)$, which depends on relative orientation $\varphi$ between complex and electric field. Introducing the rotation friction coefficient $\xi_R$ of bead-complex, the time required to escape the trap scales as
 \begin{equation}\label{time trapping}
\tau_{trap}\sim\xi_R\int_{\varphi_1}^{\varphi_2}\frac{d\varphi}{\Gamma_{el}(\varphi)}
\end{equation}
where angles $\varphi_1$ and $\varphi_2$ are respectively the orientation of complex at the beginning and the end of trapping event, cf figure ~\ref{figure1}c. It will be checked in the next section that the precise choice of these angles is not that crucial to obtain qualitative informations. Moreover, the estimation of free and trapping average time presented here are sufficient to address the questions of position- and geometry- dependence of electrophoretic mobility mentioned in the introduction. Precise formulation of trapping events is beyond the scope of this work. Simulation works with model gels (cubic arrangement of fibers) might help to unravel the details of such collision events \citep{Allison_etal02}.

\section*{Results and discussion}
\label{results}
\subsection*{Migration in pure buffer}

The application of Eq.~\ref{mu principal} for the standard geometry and conditions as defined in Appendix B is shown in figure ~\ref{figure2}. For a given position of the bead-complex, \textit{i.e.} a given length for one of the arms, the mobility oscillates as function of the relative orientation of electric field and bead-complex. This mainly reflects the anisotropy of bead-complex friction coefficient (datas not shown). As it is clearly demonstrated in figure ~\ref{figure2}, two different positions of bead-complex ($x=1$ end-position and $x=9$ middle-position) are associated with different angular average mobilities and oscillation amplitudes. A striking result is that in pure buffer, we predict slightly larger average mobility for middle-position than end-position bead-complex. Checking for both the electrophoresis and nucleosome litterature, we did not find any experimental study of capillary electrophoresis of positioned mononucleosomes, and therefore this simple prediction has not being addressed yet. 

This result in pure buffer has to be contrasted with the well-known experimental results obtained many times in gel, which show precisely the opposite: end-position nucleosomes are faster than middle-position ones during native gel electrophoresis. This discrepancy comes from the direct influence of the gel on the migration process: the porous structure of the gel provides an orientation constraint such that optimal orientation during the migration is favored rather than uniform angular average. As a consequence, our model predicts under such conditions that end-position nucleosomes move faster than middle-position ones in agreement with the experiments (cf thick dashed lines in figure ~\ref{figure2}). For interpolating positions of the nucleosome, the electrophoretic mobility changes gradually between the extreme positions. Notice that higher values of electric field might lead to orientation as well, due to the alignment of net dipole of the nucleosomes with electric field. The amplitude of positioning effect on the mobility is shown in the inset of figure ~\ref{figure2}. For the sake of simplicity,  we chosed the optimal orientation of nucleosomes for computing the positioning curve of mobility, therefore neglecting fluctuations around this optimal orientation. The precise non-uniform distribution of orientations might be quite sensitive to the gel model. A qualitative comparison of the amplitude computed with optimal orientation  with respect to experimental amplitude under similar conditions shows that the predicted amplitude of positioning effect is much weaker. Indeed ratio between middle-position and end-position mobility can be as small as 0.4 for particular conditions (see figure ~\ref{figure pennings} below for longer DNA) \citep{Meersseman_etal92}. In the next subsections, we vary different parameters of the model in order to scan the range of accessible amplitudes, and check whether the continuous model is able to reproduce the experimental range somehow.

\subsection*{Range of model parameters}

There are at least three parameters in the model than can be tuned in order to match experimental conditions: the hydrodynamic screening lengths due respectively to gel and ionic buffer, and the effective charge of nucleosome core. For all the results presented in this section, optimal orientation of nucleosomes during migration has been chosen in order to calculate the electrophoretic mobility.
The hydrodynamic screening due to the gel depends mainly on gel concentration for a fixed composition. As it is seen in figure ~\ref{figure3}, the screening length has barely no influence for sizes larger than the bead-complex itself. For smaller screening length, the positioning effect is reduced. This comes from the fact that as the range of hydrodynamic interactions decreases, the influence of position on the electrophoretic mobility decreases as well: for asymptotically very short range hydrodynamic interactions (of order of DNA bead size), the hydrodynamic influence of bead-complex arms on the core is roughly the same whathever the respective length of the arms. Notice that under standard conditions defined in the Appendix B, short range electrostatic screening is always present. The results of Pennings \textit{et al.} \citep{Pennings_etal92} can be precisely interpreted as the effect of hydrodynamic screening: they observed that the positioning effect is lost when migration is performed in glycerol, therefore when  the hydrodynamic screening is increased (without trapping).

Coming back to the situation of no hydrodynamic screening due to the gel, it is possible to observe the influence of electrostatic screening alone on the electrophoretic mobility. As it is demonstrated in figure ~\ref{figure3}, the positioning effect increases strongly as the ionic strength of the buffer is increased or equivalently as the Debye screening length is decreased. Note that the conformation of the nucleosome was artificially kept fixed under ionic strength variation, in order to address specifically the dynamic role of salt ions. 
The subtle interplay  between electrostatics and hydrodynamics provide therefore a way of modulating the amplitude of positioning effect in pure buffer.

Finally, the effective net charge of the core can also be considered as an adjustable parameter: it is difficult to assign such a value purely from theoretical considerations, because this net charge depends on many intricated features like the state of charge of the protein octamer, the ionic strength of buffer and the possible counterion condensation. What is precisely known from experiments is that the net charge of the nucleosome core is negative, the DNA overcharging the basic charge of the protein octamer. Increasing the net charge of the nucleosome core reduces the positioning effect, as can be seen from figure ~\ref{figure3}.This reflects the role of the core as the main driving force for migration, the difference in friction contribution between middle-position and end-position becoming less important as the net charge is increased.

A partial conclusion drawn from this scan of model parameters is that continuous electro-hydrodynamic model is able to describe the positioning effect on the electrophoretic mobility in a qualitative way, but not in a \textit{quantitative} way. As it will be shown in section \ref{trapping}, trapping events are more likely to be responsible for the experimentally observed amplitude of the positioning effect.

\subsection*{Bead-complex opening}
\label{bead-complex opening}

Still working within the framework of continuous electro-hydrodynamic model, it is possible to investigate the influence of bead-complex geometry on the electrophoretic mobility. In particular, we vary in this section the opening angle of the bead-complex, \textit{i.e.} the amount of DNA wrapped in the nucleosome core, mimicking different nucleosome conformations either due to incomplete formation of the histone octamer or to the presence of histone variants.

During the numerical calculation, we take into account the fact that a reduced DNA complexed length within the nucleosome reduces effectively its core net charge. For the sake of simplicity, we assume that the net charge of nucleosome core scales linearly with the complexed length of DNA.
The results are presented in figure ~\ref{figure4} for different opening angles $\theta$ as function of bead-complex orientation with respect to the electric field for middle-position nucleosomes. The discrete values of $\theta$ were chosen such that the number of beads in the arms is always an even number.   
For each opening angle, the mobility oscillates. This representation highlights the different amplitudes and relative phases of these oscillations.
Due to the gradual opening of the nucleosome as $\theta$ increases from 0$^{\circ}$ (two superhelical turn of DNA) to 360$^{\circ}$ (one superhelical turn of DNA), the optimal orientation during migration, as imposed by the gel, switches between two values (90$^{\circ}$ and 0$^{\circ}$). The net result for the predicted mobility at optimal orientation is shown in the inset of figure ~\ref{figure4}: the mobility first decreases as function of opening angle, and increases slightly for $\theta\simeq 360^{\circ}$.

These results only indicate qualitative trends for the opening angle influence for two reasons: the first one is that the variation of nucleosome core net charge might be nonlinear with respect to the DNA wrapped length due to the ion condensation phenomena. The second reason is related to histone octamer stabilization by the DNA. In a solution under physiological conditions made by the four different histones, there are mainly two populations: H3-H4 tetramer and H2A-H2B dimers, but almost no octamer that are not stable without DNA around it. This means that if less DNA is wrapped around the nucleosome core, one or two H2A-H2B dimers might be lost, and therefore the net charge of the core might be changed dramatically, as opposed to the linear variation assumed for the calculation.
One way of addressing these two points is to test for different decreasing relation $Z_{core}$ vs $\theta$. The result of such additional calculations not shown here is that the mobility is still a decreasing function of opening angle for a large range of angles: the observations presented in this subsection are quite robust with respect to core net charge variation.

\subsection*{Trapping}
\label{trapping}
In this section, we provide a qualitative estimate of electrophoretic mobility according to the trapping-untrapping scenario proposed in a previous section about specific gel effects. A more rigorous geometrical calculation is proposed in Appendix C of this work, by evaluating the trap escape time due to electrostatic torque. The results of two approaches are consistent with each others, and therefore are thought to provide a correct estimate of position-dependent mobility.

As it was previously discussed, the leading order specific gel effect influencing nucleosome migration is the occurence of frequent collisions with gel fibers. This is described approximately by introduction of two characteristic times (cf Eq. ~\ref{trap}): the free motion time and the trapping time. The former depends mainly on the gel concentration, while the latter is closely related to the bead-complex conformation. Within this context, it is straightforward to interpret the fast migration of end-position nucleosome as compared to middle-position nucleosomes: the latter adopt more likely kinked configurations and their velocity are therefore strongly reduced through trapping, while the former adopt ``tadpole''-like configurations and they have smaller probability to be trapped. 

Similarly, different opening angle lead to different trapping time. The limiting values are $\theta=0$ (two superhelical turns) with almost no trapping due to the absence of kink in the conformation, and $\theta=\pi$ (1.5 superhelical turn) with high trapping time due to a 180° kink.

In order to make a simple functional prediction for the electrophoretic mobility, one can expand the trapping time to the second order in nucleosome position or arm length $x$. Taking into account positioning symmetry considerations, this time is rewritten
\begin{equation}
\tau_{trap}\simeq\frac{A(\theta)}{E}\, x(L-x)
\end{equation}
where $L$ is the total length of nucleosome arms and $A(\theta)$ depends mainly on nucleosome net charge, rotational friction and geometry through variable $\theta$. In the previous equation, it is implicitely assumed that end-position nucleosomes are not trapped at all. As a consequence, we propose the following prediction concerning the position-dependent electrophoretic mobility in gel of a mononucleosome
\begin{equation}
\label{mu trap}\mu=\frac{\mu_0(x)}{1+\frac{d^2A(\theta)\mu_0(x)}{c_g}x(L-x)}
\end{equation}
where $\mu_0(x)$ is the mobility in pure buffer, as calculated in previous sections. Since variation of mobility with position in pure buffer is less than 6-7 $\%$ whatever the range of realistic parameters tested in this study, one can use constant $\mu_0$ in a first approximation in order to use our prediction to interpret experimental datas. Equation ~\ref{mu trap} provides therefore a \textit{simple semi-empirical} formula that is designed to rationalize experimental datas of position-dependent electrophoretic mobility. The application of such a formula for constant $\mu_0$ to the experimental datas of Meersseman \textit{et al.} \citep{Meersseman_etal92} on mononucleosome positioning on twofold repeat of 5S rDNA sequence (total length=414bp) is shown in figure ~\ref{figure pennings}. A reasonable agreement between model and experimental datas is found. However, the correct interpretation of the fitting parameter would require additional systematic experimental datas. Therefore we do not pursue further the analysis of fit parameters. A clear conclusion that can nevertheless be drawn at this level of analysis is that trapping events are mainly determining the amplitude of positioning effects, and therefore they cannot be neglected in the theoretical interpretations of datas.

\subsection*{Applications and limitations} 
\label{applications limitations}
In this work, we presented a model that can be used in order to interpret the position-dependent electrophoretic mobility of mononucleosomes. In a first step we computed the mobility in pure buffer, and then we took into account specific effects associated to the migration in gel.
The theory describing the migration in pure buffer is based on continuous electro-hydrodynamic description as applied to a coarse-grained model of nucleosome made of beads of various radii and net charges. One of the main simplifying assumption used in order to derive an explicit expression out of the compact formula Eq.~\ref{mu principal} is the Debye-Huckel approximation that leads to short-ranged Rotne-Prager-like tensors (cf eq.~\ref{RP screened1}). 
At first sight, it might appear very naive to apply it for the electrostatic potential around such a highly charged object like the nucleosome \citep{Schiessel03}. However it allows to hide the effect of complicated features like counterion condensation in the effective charges of beads in the coarse-grained model. Moreover, the exponential decay of electrostatic potential at longer range is also expected from more rigorous approaches. In the results presented in previous sections, we used the same effective charge for each DNA-bead. An improvement of the model would be to take into account for inhomogeneous counterion condensation on the DNA-beads forming the arms and therefore non-uniform DNA-bead charges, since it is known that electrostatic potential along a finite size polyelectrolyte is not constant \citep{Castelnovo_etal00,Allison94}. The order of magnitude of changes of positioning effect on electrophoretic mobility due to inhomogeneous counterion condensation is similar to the one obtained by varying the net charge of the core $Z_{core}$ (unpublished results).

Our numerical calculations shows that the positioning effect in pure buffer and in gels are opposite and different in magnitude: end-position nucleosomes move faster (compared to middle-position nucleosomes) in gels, while they are slower in pure buffer. This is mainly explained by the orientation of the nucleosome during gel migration as opposed to an orientationally-averaged migration during pure buffer electrophoresis at low electric field. However the small amplitude of position-dependent mobility in pure buffer might be difficult to measure experimentally. A personal interpretation of our results is that capillary electrophoresis, although not used systematically for protein-DNA complexes characterization, might bring new information on these systems, because both analytical and numerical hydrodynamic models used to interpret the datas are becoming more precise and powerful (see for example \citep{Allison_etal04}).

Using the continuous electro-hydrodynamic model, it is possible to investigate the influence of nucleosome geometry on the value of electrophoretic mobility, as it is described in previous section. The main result is that the mobility of middle-position nucleosome decreases as the DNA length wrapped in the core is decreasing. Focusing on incomplete nucleosome characterization, our model would predict that the fastest specie in pure buffer is the octamer, then the hexamer and finally the tetramer. Here again the experimental results are different in gels: the fastest are hexamer, then octamer and finally tetramer. The mobility of the octamer relative to the other specie is not correctly evaluated through the continuous model. This discrepancy can be partially interpreted as a specific effect of the gel using the trapping model defined in previous section: although the precise DNA wrapped length of hexamer (between 1 and 1.5 superhelical turn )and tetramer (roughly one superhelical turn) is not known, one can speculate that the kink formed by the two DNA arms is less important than for the octamer (90$^{\circ}$ kink or  1.75 superhelical turn). Therefore the octamer is more sensitive to trapping mechanism, and its mobility is further reduced. As a consequence the octamer will not be the fastest specie anymore in gel. It can be either the second fastest or the third fastest. The former situation corresponds to the experimental observation. However at the level of the present description, we can not discriminate between the two cases. However, it is clear that the role of core net charge will be important.

Using the trapping model, we propose a \textit{semi-empirical} formula Eq.~\ref{mu trap} in order to predict the electrophoretic mobility of positioned nucleosomes. The application of such a formula to experimental datas of Meersseman  \textit{et al.} seems a reasonable guess \citep{Meersseman_etal92}. This leads us to propose the following method in order to determine unknown positioning within a single gel run. The idea is that for this single gel run, one should have in a lane two different known positions (middle and end position for example) on a well-known sequence in order to provide a ladder. Then the two-parameters formula Eq.~\ref{mu trap} (at constant $\mu_0$) can be used to get the unknown positions on another sequence in different lanes, provided that the DNA length on which mononucleosomes are constructed is the same.

As a conclusion, the main gain of the approach presented in this work is that it provides a rigorous framework for the understanding of position-dependent electrophoretic mobility of mononucleosomes. Moreover, the influence of different experimental parameters can be qualitatively predicted. This work may serve as a guideline for more thorough studies of electrophoresis of rigid molecular complexes. We are currently developping similar models in order to investigate more thoroughly the dependence of electrophoretic mobility of curved DNA on bent angle and position \citep{Castelnovo_06}.

\textbf{Acknowledgments}--
Fruitful discussions with H. Menoni, D. Anguelov and P. Bouvet are gratefully acknowledged. The authors thank S.A. Allison for useful comments on this work.

\section*{Appendix A: friction coefficients for screened hydrodynamics}
In order to calculate the friction coefficient of a single bead into a fluid, it is necessary to solve the flow and pressure profile around this particle. In the case where the hydrodynamics is screened either because of electrostatic screening or because of the gel, the calculation of the friction coefficient can still be done analytically. Although the flow profiles are similar in the two situations, the friction coefficient are different due to different pressure field \citep{Long_etal01}. The reader is referred to the works of Russel \textit{et al.} \citep{Russel_etal89} or Stigter \citep{Stigter00} for futher details on the derivation of the friction coefficients. The result goes as follows for electrostatic screening:
\begin{eqnarray}
\xi_{el} & = & 6\pi\eta \sigma_1 (1+\kappa_D\sigma_1)/[1+\frac{1}{16}(\kappa_D\sigma_1)^2-\frac{5}{48}(\kappa_D\sigma_1)^3-\frac{1}{96}(\kappa_D\sigma_1)^4\nonumber\\
& & +\frac{1}{96}(\kappa_D\sigma_1)^5+\left\{\frac{1}{8}(\kappa_D\sigma_1)^4-\frac{1}{96}(\kappa_D\sigma_1)^6\right\}e^{\kappa_D\sigma_1}E_1(\kappa_D\sigma_1)]
\end{eqnarray}

with the exponential integral $E_1(x)=\int_x^{\infty}\frac{e^{-u}du}{u}$. In the case of gel screening, the friction coefficient simply reads:
\begin{equation}
\xi_{gel} =6\pi\eta\sigma_1\left(1+\kappa_g\sigma_1+\frac{1}{9}(\kappa_g\sigma_1)^2\right)
\end{equation}

\section*{Appendix B: geometry of the bead-complex}
In this appendix, we describe the geometry of the complex that mimicks electro-hydrodynamics of positioned nucleosome. This complex is shown in figure ~\ref{figure1}. According to structural datas available on mononucleosomes \citep{Luger_etal97}, DNA is wrapped on a superhelical path around an octamer of histones. In the case of positioned nucleosome, DNA arms entering and exiting the nucleosome core are also present. Their conformations depend mainly on ionic strength of the buffer (see for example \citep{Kunze_etal00}). In the present work, we assume for the sake of simplicity that DNA outside the nucleosome core is following a straight path, whose direction is given by the tangent path of the last bead in the complex core . This is justified by both the rigidity of DNA backbone (persistence length of roughly 150 base pairs) for such small non-wrapped lengths of nucleosomes considered ($<100bp$) and by the physiological ionic strength that effectively screens electrostatic interactions beyond 1 to a few nanometers. 

Due to the level of description of both hydrodynamic and electrostatic interactions in this work, the nucleosome core (histone octamer and 147 base pairs of DNA) is represented by a single bead with an effective radius $R_{core}$ and effective net charge $Z_{core}$. Indeed, we use Rotne-Prager tensors as well as Debye-Huckel approach for interactions. These expression are correct for large separations, as long as effective radii and charges are taken into account. Moreover in the case of electrostatics, subtle effects like net charge of protein and counterion condensation are taken into account by the appropriate choice of $Z_{core}$.

Protruding from the central core bead, DNA arms are represented by two linear arrays of smaller beads. Due to the natural anisotropy of a base pairs of radius $r_{bp\perp}=1nm$ and height $r_{bp||}=0.34nm$, a single ``DNA'' bead embeds several base pairs. For a given number of base pairs $N_{bp}$, the number of beads in the two arms is given by  

\begin{equation}
N_{bead}= \frac{(N_{bp}-\frac{147(4\pi-\theta)}{(1.75)2\pi}-1)r_{bp||}}{2r_{bp\perp}}+1
\end{equation}

It has been implicitely assumed in the previous equation that in the reference nucleosome 147 base pairs of DNA are \textit{exactly} wrapped into 1.75 superhelical turns. 
This formula allows to calculate the number of beads in the arms for different opening angle $\theta$, cf figure ~\ref{figure1}.
The first bead of each arms is tangent to the central core bead, and is located at the coordinate of the last base pair of the nucleosome core (base pairs 1 and 147). Similar results are obtained if slightly different matching conditions are used.

The choice of values for the bead-complex effective parameters is made mainly following values used in related brownian dynamics simulations by Beard and Schlick \citep{Beard_etal01}. In the main part of this work, we refer ``standard geometry and conditions'' to the following set of parameters

\begin{center}
\begin{equation}\begin{array}{ccc}
N_{bp}=250bp & R_{core}=5nm & R_{bead}=1nm \\
\theta=\frac{\pi}{2} & Z_{core}=200 & Z_{bead}=8.3\\
\kappa_D^{-1}=1.35nm & \mathrm{- \,no \, gel \, screening-} & \mathrm{-no \, trapping-} 
\end{array}\end{equation}
\end{center}

\section*{Appendix C: estimation of untrapping time for crossed configuration}

Although, the precise collision scenario is not known, one might anticipate that the longest time (which is the relevant time to estimate for the mobility calculation) is associated with the rotation of bead-complex around gel fibers. This rotation is driven by the electric field that exerts a torque on the complex.

An estimation of this torque is simply made in the case of planar configuration of the complex (cf figure ~\ref{figure5}). The result reads 
\begin{eqnarray}
\frac{T_{EH}}{qE} & = & \sin \varphi \bigg[ \sqrt{1+\tan^2\frac{\theta}{2}}\left(\frac{Z_c}{q}(R_{core}+R_g+2R_{bead})+(L_1+L_2)(R_g+R_{bead})\right) \nonumber\\
 & & -\sin \frac{\theta}{2} \left(\frac{L_1^2+L_2^2-4(R_{core}+R_{bead})^2\tan^2\frac{\theta}{2}}{2}\right) \bigg]\nonumber\\
 & & + \cos \varphi \bigg[\cos\frac{\theta}{2}\left(\frac{L_2^2-L_1^2}{2}\right)\bigg]
\end{eqnarray} 
where $R_g=d/2$ is the gel fiber radius. The charge density of DNA is $q=Z_{bead}/2R_{bead}$. The length of the two arms are $L_1$ and $L_2$.

According to Eq. ~\ref{time trapping}, the trapping time is mainly determined by integrating  the inverse of electrostatic torque between two angles $\varphi_1$ and $\varphi_2$ that represent respectively the initial and final orientations of the bead-complex during the collision. Instead of calculating such integrals, and eventually averaging over initial and final angles,  we plotted the inverse of the torque for various arm lengths at fixed opening angle $\theta$ in figure ~\ref{figure5}b since we are mainly interested in the way trapping time changes with nucleosome position: the main result is that for any couple of reasonable angles $\varphi_1$ and $\varphi_2$, the trapping time increases going from end-position to middle-position nucleosomes since curves sit on top of each others without crossing. As a consequence, this simple argument shows qualitatively that end-position nucleosome will migrate faster than middle-position ones in a scenario where trapping determines the dynamics. Similarly, plotting the inverse torque for various opening angle at fixed arm lengths shows that trapping time increases with opening angle in the range $\theta=[0,\pi]$, and therefore the mobility decreases with the opening angle in the same range, in qualitative aggreement with the results of continuous electro-hydrodynamics model.

\bibliography{biblio}

\clearpage
\section*{Figure Legends}
\subsubsection*{Figure~\ref{figure1}.}
Gel electrophoresis of bead-complex: \textit{(a)} geometry of the bead-complex and definition of parameters $\varphi$ the orientation between complex and electric field, $\theta$ the opening angle of the complex and $x$ the length of one arm of the complex; the black and grey beads represent respectively DNA and nucleosome core (DNA+histones); the electric field direction is shown by an arrow.  \textit{(b)} Illustration of orientation constraint on bead-complex migration in a gel. Black squares represent cross-section of gel fibers. \textit{(c)} Illustration of two-state motion due to trapping-untrapping events.
\subsection*{Figure~\ref{figure2}.}
Electrophoretic mobility of bead-complex in pure buffer. \textit{Main panel}: orientation dependence with respect to eletric field of end-position (squares) and middle-position (circles) bead-complex. Thick grey and black lines represent respectively uniform angular average mobility of end-position and middle-position bead-complex. Dashed thick grey and black lines represent respectively most favorable mobility value (cf orientation constraint due to the gel) for end-position and middle-position bead-complex. \textit{Inset}: Influence of bead-complex positioning on relative mobility ratio for most favorable orientation. $x$ is the number of bead in one of the arms of bead-complex.
\subsection*{Figure~\ref{figure3}.}
Influence of hydrodynamic screening and core particle net charge. \textit{Left panel}: Relative mobility (middle- \textit{vs} end- position) as function hydrodynamic screening length $\kappa_{g}^{-1}$ (nm) due to the gel. \textit{Center panel}: Relative mobility as function of electrostatic screening length $\kappa_D^{-1}$ (nm). \textit{Right panel}: Relative mobility as function bead-complex core net charge $Z_{core}$.
\subsection*{Figure~\ref{figure4}.}
Influence of bead-complex opening. \textit{Main panel}: Mobility in pure buffer for standard geometry as function of orientation with respect to the electric field for different opening $\theta=\pi/100,\pi/3.5,\pi/2,\pi/1.25,\pi*1.15,\pi*1.4,\pi*1.9$. \textit{Inset}: Mobility for middle-position bead-complex as function of opening angle $\theta$ for optimal orientations.
\subsection*{Figure~\ref{figure pennings}.}
Experimental mobility of Meersseman \textit{et al.} \citep{Meersseman_etal92} as function of dyad position $x_{pos}$ fitted by the prediction Eq.\ref{mu trap}. The datas were obtained on mononucleosomes constructed on a twofold repeat of 5S rDNA (total length=414bp).
\subsection*{Figure~\ref{figure5}.}
Estimation of trapping time $\tau_{trap}$. \textit{(a)} Geometry considered for the rotation of bead-complex due to electrostatic torque. \textit{(b)} Inverse of electrostatic torque as function of orientation. \textit{Upper panel}: at fixed opening angle ($\theta=\pi/2$), different curves correspond to different asymmetry of arms 40bp-60bp,30bp-70bp,20bp-80bp,10bp-90bp. \textit{Lower panel}: at fixed asymmetry of arms 40bp-60bp, different curves correspond to different opening angle $\theta=\pi/1.5,\pi/2,\pi/3,\pi/10$.

\clearpage

\clearpage
\begin{figure}
   \begin{center}
      \includegraphics*[width=14cm]{figure2}
      \caption{}
      \label{figure2}
   \end{center}
\end{figure}

\clearpage
\begin{figure}
   \begin{center}
      \includegraphics*[width=14cm]{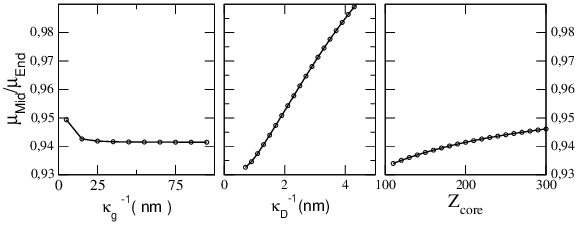}
      \caption{}
      \label{figure3}
   \end{center}
\end{figure}

\clearpage
\begin{figure}
   \begin{center}
      \includegraphics*[width=14cm]{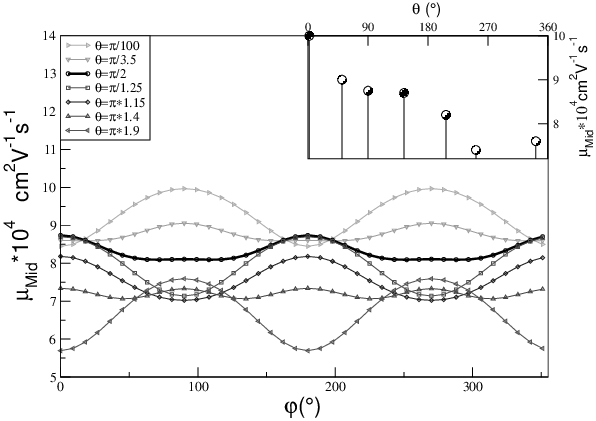}
      \caption{}
      \label{figure4}
   \end{center}
\end{figure}

\clearpage
\begin{figure}
   \begin{center}
      \includegraphics*[width=14cm]{figure5}
      \caption{}
      \label{figure pennings}
   \end{center}
\end{figure}

\clearpage
\begin{figure}
   \begin{center}
      \includegraphics*[width=14cm]{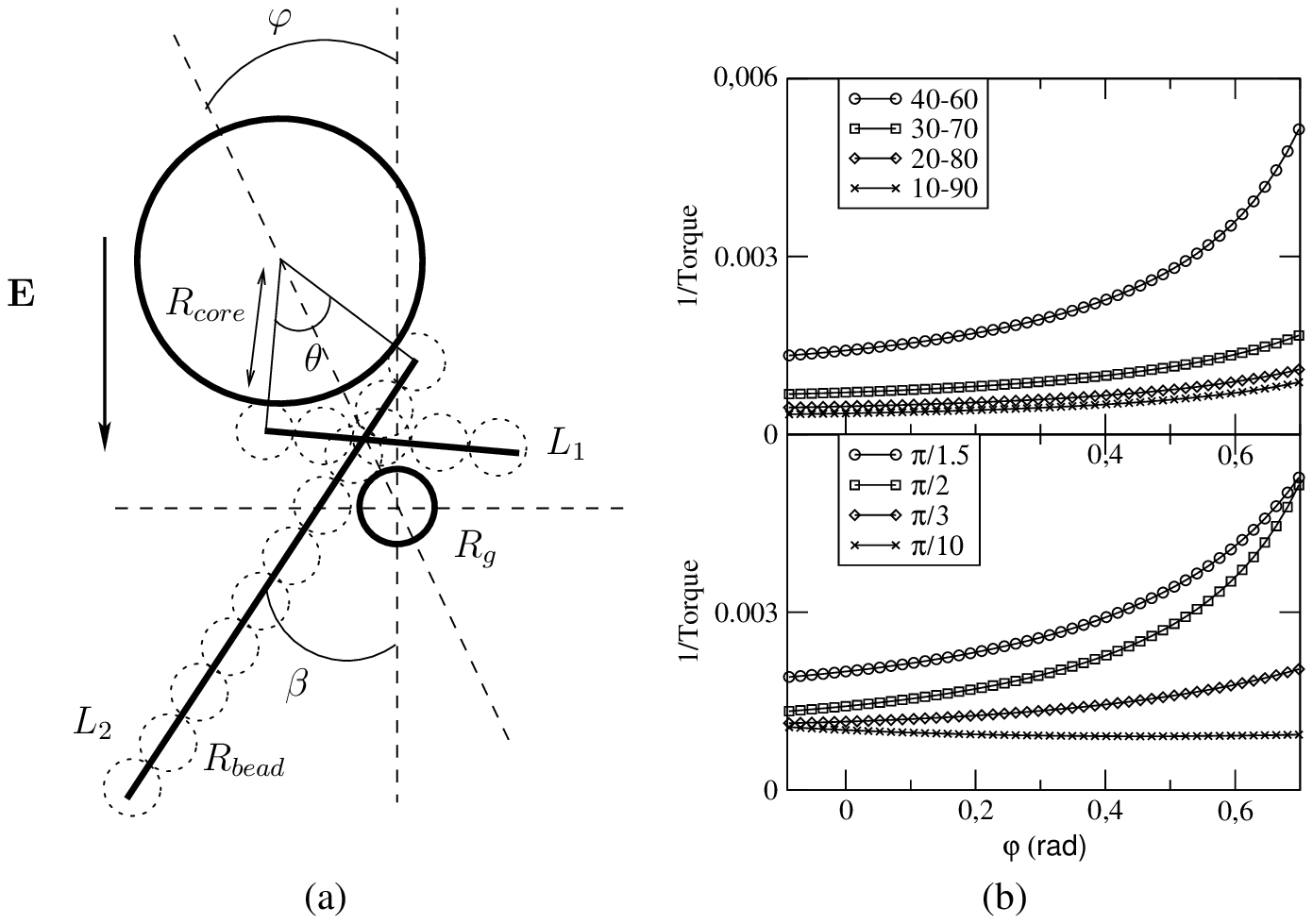}
      \caption{}
      \label{figure5}
   \end{center}
\end{figure}

\end{document}